# Observation of Neutrons with a Gadolinium Doped Water Cerenkov Detector

S. Dazeley[a,*], A. Bernstein[a], N. S. Bowden[a] and R. Svoboda[b,a]
[a]*Lawrence Livermore National Laboratory, Livermore, CA 94550, USA*
[b]*Department of Physics, University of California, Davis, CA 95616, USA*

**Abstract:**
Spontaneous and induced fission in Special Nuclear Material (SNM) such as $^{235}$U and $^{239}$Pu results in the emission of neutrons and high energy gamma-rays. The multiplicities of and time correlations between these particles are both powerful indicators of the presence of fissile material. Detectors sensitive to these signatures are consequently useful for nuclear material monitoring, search, and characterization. In this article, we demonstrate sensitivity to both high energy gamma-rays and neutrons with a water Cerenkov based detector. Electrons in the detector medium, scattered by gamma-ray interactions, are detected by their Cerenkov light emission. Sensitivity to neutrons is enhanced by the addition of a gadolinium compound to the water in low concentrations. Cerenkov light is similarly produced by an 8 MeV gamma-ray cascade following neutron capture on the gadolinium. The large solid angle coverage and high intrinsic efficiency of this detection approach can provide robust and low cost neutron and gamma-ray detection with a single device.

**Introduction:**
Many nuclear material monitoring and surveillance applications would benefit from more affordable, deployable, and environmentally benign gamma-ray and neutron counters. Long standing international nonproliferation treaties call for the tracking of nuclear materials in civil nuclear fuel cycles. More recently, there has been increased interest in the use of radiation-based screening and interrogation methods to discover illicit SNM in cargo. In response, a variety of radiation portal monitors (RPM) have been developed to detect the neutrons and gamma-rays emitted by SNM, whether spontaneously or enhanced with interrogating beams.

Practical limitations, such as the need to monitor large cargo containers quickly in order to not impede legitimate commerce, require the use of large, highly efficient detection systems. Because the signature of SNM, crudely speaking, consists of several neutrons and gamma rays emitted isotropically and in close time proximity, large solid angle coverage is imperative. This recognition has led to the use of large organic plastic or liquid scintillator detectors [1 - 4]. These devices have a relatively poor spectral response. As a result, these methods often rely on detecting an increase in the rate of incoming particles compared to background, rather than the analysis of the particle energy spectrum. Even without spectroscopy, rate-based approaches, such as the 'nuclear car wash' interrogation scheme [4], have been shown to be effective.

---

[*] Corresponding author: Lawrence Livermore National Laboratory, 7000 East Ave, L-211, Livermore, CA, 94550. ph: 925 423 4792 *Email address: dazeley2@llnl.gov*

Well shielded SNM is difficult to detect, or may escape detection entirely, especially if the detector is sensitive to only gammas or only neutrons. A system sensitive to both gammas and neutrons is more difficult to effectively shield against, since effective gamma ray shielding requires high Z material, while that for neutrons requires low Z material. If the dominant gamma ray interaction in the detection medium is Compton scattering, a useful device must also be sufficiently thick to contain a significant fraction of the fission gamma-ray energy.

Although organic scintillators are attractive in many respects, they have several drawbacks. Plastic organic scintillators are expensive to build in large volumes, and difficult to dope with neutron capturing agents like Gd or $^6$Li. Liquid scintillator is often toxic and highly flammable. Given these limitations water Cerenkov based detectors offer an interesting alternative. Water is non-toxic, non-flammable and inexpensive, and retains these properties even after doping with Gadolinium or other neutron absorbing compounds. For example, the Super-K and SNO experiments [5,6] have shown that the Cerenkov process can generate enough photons to permit detection of gamma-rays with an energy greater than about 3 to 4 MeV - or neutron captures on chlorine - so long as the photocathode coverage is high (~40%). Neutron capture on chlorine (like gadolinium) produces an 8 MeV gamma-ray cascade. SNO observed that the light output was approximately equivalent to a 6 MeV electron. As demonstrated here, the addition of highly reflective white walls makes it possible to detect gammas of a few MeV or more with a much smaller photocathode coverage - only about 10%.

Weapons grade Plutonium (WGPu) and highly enriched Uranium (HEU) are copious emitters of low energy (sub MeV) gamma-rays. However, if a source is located within a moderate amount of shielding, this gamma-ray signature is suppressed and difficult to detect. WGPu contains mostly $^{239}$Pu, but also a few percent of $^{240}$Pu, which spontaneously fissions, emitting approximately three neutrons on average per fission. The multiplicity (number) distribution of emitted neutrons and gamma rays are roughly Poissonian, while the time correlations between successive neutrons and gamma-rays from the source are non-Poissonian. A large solid angle water Cerenkov detector sensitive to neutrons could detect such a source, either by registering an increase in the raw count rate or by observing a time correlation between sets of neutrons and gammas emitted in a single fission reaction or in fission chains. Methods based on interrogation with neutron or gamma beams can also benefit from this type of mixed gamma-ray/neutron detector. For example, HEU can be induced to undergo fission due to bombardment by low energy neutrons [7] or by high energy photonuclear induced fission [8]. After fission is successfully induced, the resulting unstable nuclei can produce so-called 'β-delayed' gamma-rays above 3 MeV over several tens of seconds. These high energy gamma-rays should be detectable in a water Cerenkov detector. Additionally, if such a detector were sensitive to neutron capture, it may be possible to observe time correlations between fission gammas and/or neutrons.

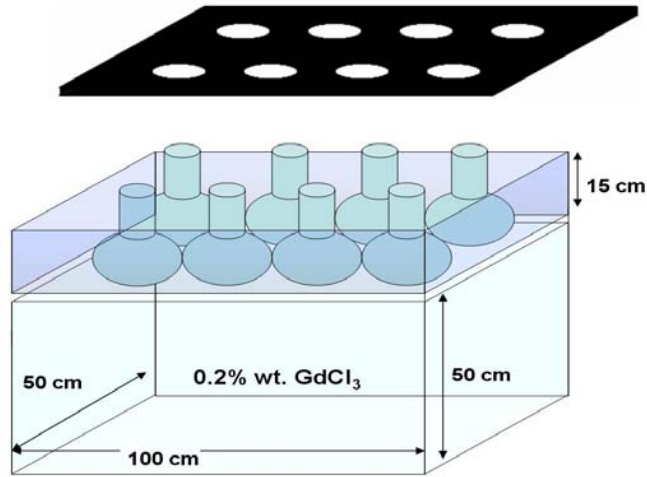

**Figure 1: Schematic design of water Cerenkov detector (see text for description)**

**Detector description:**
Our water Cerenkov detector consists of two separate acrylic tanks. A small tank sits on top of a large tank. An O-ring between the two tanks seals the volume of the lower tank. This lower tank (1m x .5m x .5m) contains ultra pure, sterilized water doped with 0.2% $GdCl_3$. This tank forms the main 250 liter active target volume of the detector. It was fitted with a small expansion volume and airlock so that the target remains full (i.e. optically coupled to the top tank) and closed to outside air despite ambient air pressure variations. The upper tank contains ultra pure, sterilized and deionized water without the Gd dopant, and eight downward facing 8 inch ETL 9354kb PMTs. The ETL PMTs have a relatively high quantum efficiency at short wavelengths (~ 30%), which is particularly advantageous for Cerenkov light detection. The PMTs face downwards into the lower tank and are individually shielded from magnetic field effects by 8 inch diameter cylinders of mu-metal. The PMT and target volumes were separated to prevent exposure of the $GdCl_3$ doped water to the mu-metal surface -- iron has been shown to react with $GdCl_3$ in water, reducing the water clarity over time [9].

Surrounding the two water tanks is a five sided muon veto. Figure 1 shows a schematic of the inner water detector, showing the two acrylic tanks and the PMTs inside. Two flanges are shown, the top one forms the lid of the upper small tank. The lower flange forms the barrier between the small upper tank and the lower target tank. The bases of the eight PMTs are shown penetrating the top of the detector. The following features were built in to the design in order to maximize the detection efficiency of the relatively small number of photons produced by the Cerenkov process:
1) Maximized light reflection within the target region. The acrylic tanks were constructed from UV transmitting acrylic. Reflection was achieved by a combination of total internal reflection off the acrylic/air boundary and UV reflective 1073B Tyvek [10,11].
2) Photocathode coverage of 10%, eight 8 inch PMTs
3) Individual mu-metal cylinders to shield the PMTs from external magnetic fields (not shown)

Even with the addition of only 0.2% $GdCl_3$, the remarkably high neutron capture cross-section of Gadolinium at thermal energies (49,000 barns) reduces the mean neutron capture time to about 30 µs from the 200 µs typical of pure water (neutron capture on hydrogen). In addition, neutron capture on gadolinium results in a gamma cascade with total energy 7.9 MeV for $^{157}Gd$ and 8.5 MeV for $^{155}Gd$, resulting in an increased probability of detection.

**Results:**
We tested the ability of the device to detect both high energy gammas and neutron capture. A fission source ($^{252}Cf$, 55 µCi or 2.4 x $10^5$ neutrons/s) was placed on the concrete floor of the testing laboratory approximately 1 meter from the detector behind a two inch lead wall as shown in Figure 2. The summed PMT response from all eight PMTs is shown in Figure 3. The raw event rate increased from 700 Hz to 7 kHz due to the presence of the source alone. Despite the poor energy resolution characteristic to the technique, it is obvious from the figure that the presence of a $^{252}Cf$ source changes the spectral shape, increasing the high energy (>10 photo-electrons) component relative to the background. The figure also indicates that if a lower background signal is required, some improvement in signal to noise may be possible with a modest energy cut.

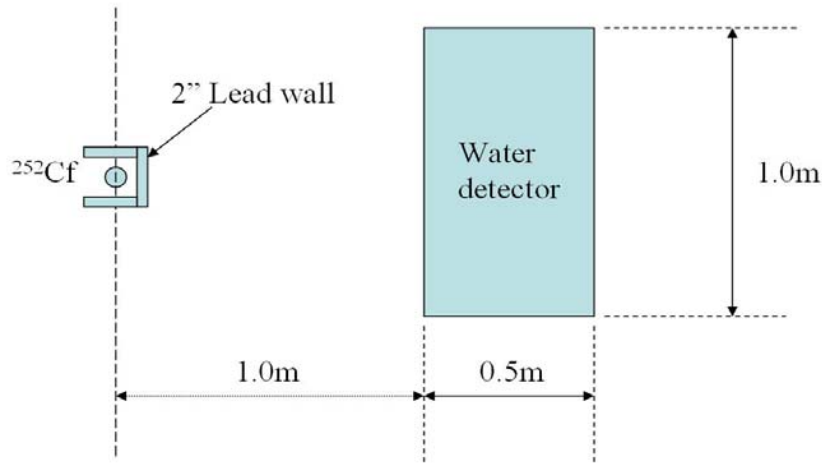

**Figure 2: Representation of the relative positions of the detector and the neutron fission source, as seen from above**

To test whether some of the signal that results from the $^{252}Cf$ source is due to neutron capture, we plot the inter-event time between consecutive events in Figure 4. Assuming Poisson statistics, the primary exponential in the presence of the $^{252}Cf$ source gives a corresponding event rate of 7 kHz. If we subtract this exponential, we are left with a correlated component at small inter-event times, which can be fit by another exponential, corresponding to a mean inter-event time of 28 µs. The 28 µs component agrees quite well with the expected mean neutron capture time in water or liquid scintillator (see [12-14]) for a gadolinium concentration of 0.1% ($GdCl_3$ concentration of 0.2%). The correlated component can be a result of either the detection of a prompt gamma ray followed by a delayed neutron capture or a prompt neutron capture followed by a delayed neutron capture (where both neutrons were emitted in the same fission event). There is also a smaller correlated neutron capture signal present when there is no $^{252}$Cf source present. This is due to spallation caused by the passage of muons near the detector and muon capture nearby, both of which may result in gammas and neutrons, or multiple neutrons due to the same muon, and hence a correlation. We claim strong evidence for neutron detection in our detector on the strength of the correlated signals in Figure 4 and the change in shape of the energy spectrum in the presence of the $^{252}$Cf in Figure 3. By employing a statistical subtraction technique, it is possible to extract the spectral shape of the neutron capture events (Fig. 3b). We extracted this by taking the spectrum of a selection of $^{252}$Cf events correlated with a prompt signal (<100 μs) and then subtracting the spectrum of a normalized selection of uncorrelated non $^{252}$Cf events. The number of non $^{252}$Cf events subtracted was calculated by fitting an exponential to all the uncorrelated events (>100 μs) in the presence of the $^{252}$Cf source and calculating the area under the line of best fit between 0 and 100 μs. This spectrum looks superficially similar to the neutron/gamma spectrum, indicating that our detector is not able to distinguish on and event by event basis between high energy fission gammas and the 8 MeV capture gamma cascade on the basis of energy alone.

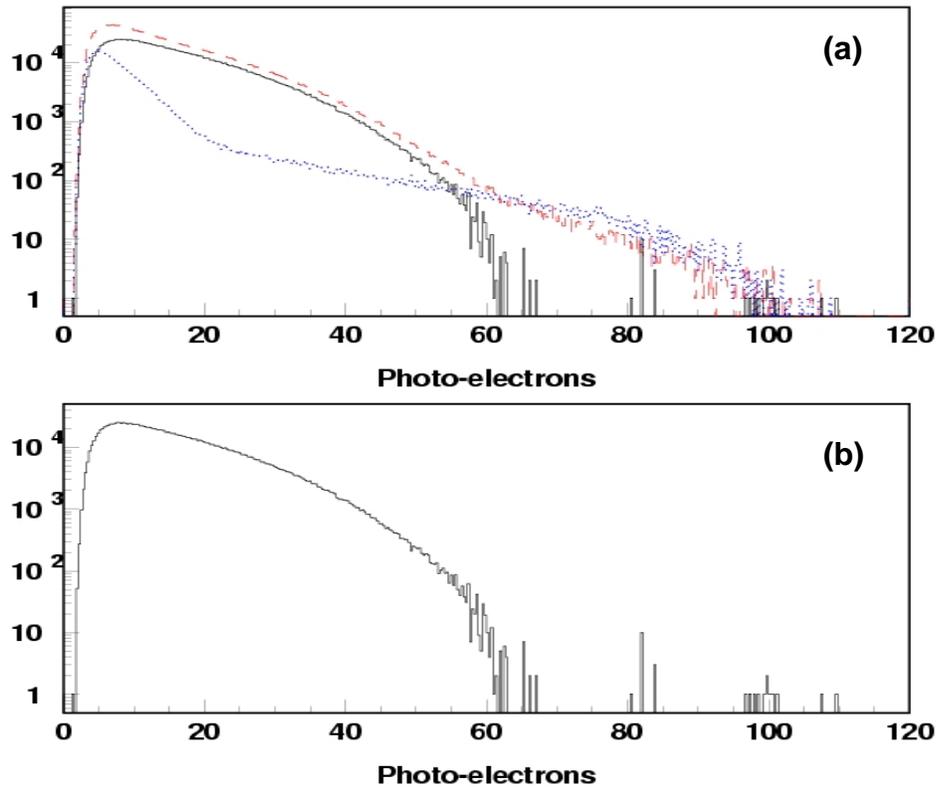

**Figure 3: The upper panel shows the summed PMT detector response spectra of our prototype with (dashed line) and without (dotted line) the presence of a $^{252}$Cf source. The difference between the two, representing the spectrum of a pure sample of neutron capture events in our prototype is shown as a solid line in both the top and bottom panels.**

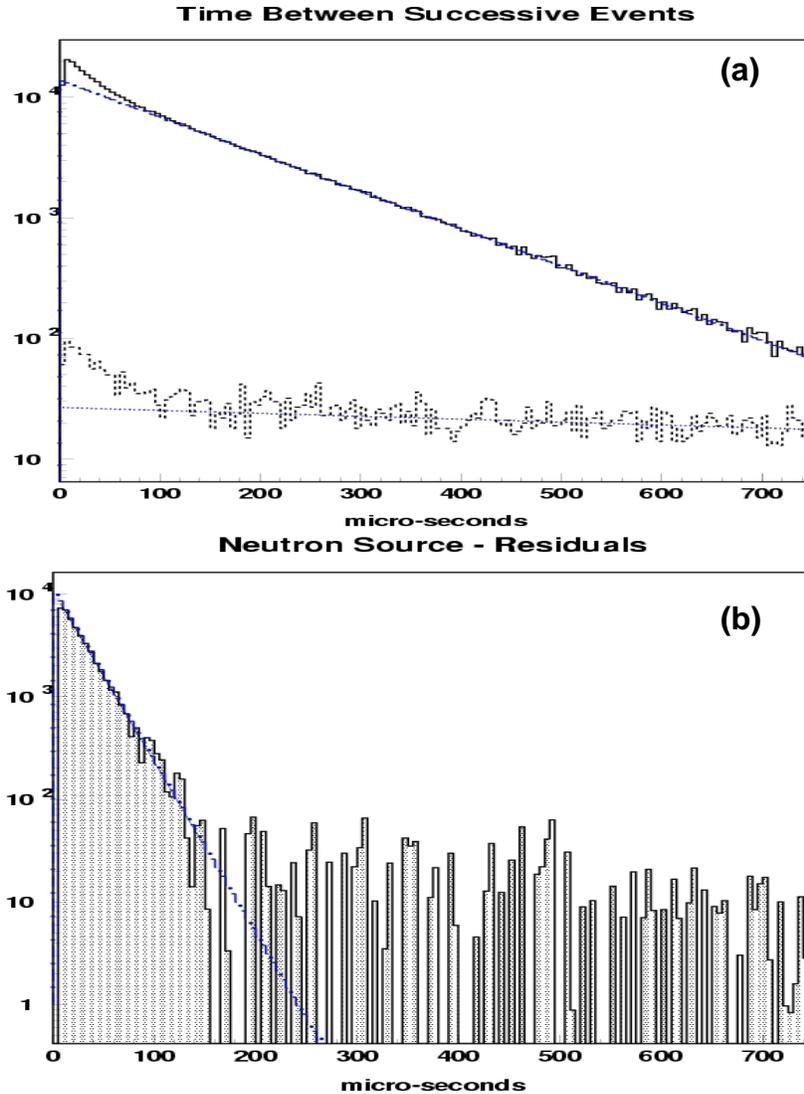

**Figure 4:** Plot of the inter-event time distribution. The upper plot (a) shows the resultant distributions with and without the presence of a $^{252}$Cf source. The source results in an increase in both the random trigger rate and the correlated trigger rate. The lower plot (b) shows the residuals that result when the random events which were fitted by an exponential curve are subtracted, leaving only the correlated component with a mean inter-event time of 28us.

**Conclusion:**
Although it has a long history of being suggested for a variety of purposes ([15,16]) Gd doping has not to our knowledge actually been used in water detectors prior to this experiment. We have designed and built a new detector which detects high energy (>~3MeV) fission gammas and thermal neutrons, using a gadolinium doped volume of water read by PMTs. Upon exposure of the detector to a $^{252}$Cf source, we observed a clear signature consistent with neutrons. We also observed a correlated signal in the inter-event time distribution, again consistent with a 28 micro-second capture time of thermalized neutrons in water doped at the 0.1% level with Gadolinium. The origin of this signal is the detection of correlated gamma-ray/neutron or neutron/neutron pairs originating from $^{252}$Cf fissions. We have operated the detector for several months with no evident diminution in its sensitivity to gamma-rays or neutrons.

**Acknowledgements:**
The authors would also like to thank John Steele for programming the trigger logic in the FPGA. This work was performed under the auspices of the US Department of Energy by Lawrence Livermore National Laboratory under contract DE-AC52-07NA27344, release number LLNL-JRNL-405708. The authors wish to thank the DOE NA-22 for their support of this project.

**References:**
[1] D. R. Slaughter, M. R. Accatino, A. Bernstein, et al., Nuclear Instruments and Methods in Physics Research B 241, 777, 2005
[2] C. E. Moss, M. W. Brener, C. L. Hollas, W. L. Myers, Nuclear Instruments and Methods in Physics Research B 241, 793, 2005
[3] J. Ely, R. Kouzes, J. Schweppe, et al., Nuclear Instruments and Methods in Physics Research A 560, 373, 2006
[4] D. R. Slaughter, M. R. Accatino, A. Bernstein, et al., Nuclear Instruments and Methods in Physics Research A 579, 349, 2007
[5] The Super-Kamiokande Collaboration, Phys.Rev. D73 112001 2006
[6] The SNO Collaboration, Phys.Rev. C72 055502 2005
[7] D. Dietrich, C. Hagmann, P. Kerr et al., Nuclear Instruments and Methods in Physics Research B 241, 826, 2005
[8] J. L. Jones, W. Y. Yoon, D. R. Norman, et al., Nuclear Instruments and Methods in Physics Research B 241, 770, 2005
[9] W. Coleman, A. Bernstein, S. Dazeley and R. Svoboda, Submitted to Nuclear Instruments and Methods in Physics Research A
[10] "Directional Reflectance Measurements(DR) on Two special sample Materials-Final Report", **SOC-R950-001-0195**, Prepared for the University of California, Irvine, School of Physical Sciences Irvine, California 92717-4675, January 1995.
[11] "Directional Reflectance (DR) Measurements on Five (5) UCI Supplied Sample Materials", **SOC-R1059-001-0396,** Prepared for the University of California, Irvine, School of Physical Sciences Irvine, California, March 1996.
[12] M. Apollonio et. al., European Phys. J. C27, 331, 2003
[13] F.Boehm et al., Phys. Rev. D64,112001, 2001
[14] A. G. Piepke, S. W. Moser, and V. M. Novikov, Nuclear Instruments and Methods in Physics Research A 432, 392, 1999
[15] J. F. Beacom, M. R. Vagins, Phys.Rev.Lett. 93, 171101, 2004
[16] A. Bernstein, T. West, V. Gupta, Science and Global Security, Volume 9 3:235:255 2001